\documentclass[twocolumn,prd,showpacs,10pt,superscriptaddress,nofootinbib]{revtex4}
\usepackage{graphicx}
\usepackage{color}
\usepackage{amsmath}
\usepackage{enumitem}
\usepackage{soul}
\usepackage{tabularx}
\usepackage[colorlinks=true,citecolor=blue]{hyperref}

\def\be{\begin{equation}}
\def\ee{\end{equation}}
\def\bea{\begin{eqnarray}}
\def\eea{\end{eqnarray}}
\def\ba{\begin{array}}
\def\ea{\end{array}}
\def\bq{{\bf q}}
\newcommand{\lsim}{\,\raise 0.4ex\hbox{$<$}\kern -0.8em\lower 0.62ex\hbox{$\sim$}\,}
\newcommand{\gsim}{\,\raise 0.4ex\hbox{$>$}\kern -0.7em\lower 0.62ex\hbox{$\sim$}\,}
\newcommand{\mpl}{{M_\mathrm{Pl}^2}}

\newcolumntype{L}[1]{>{\hsize=#1\hsize\raggedright\arraybackslash}X}%
\newcolumntype{R}[1]{>{\hsize=#1\hsize\raggedleft\arraybackslash}X}%
\newcolumntype{C}[1]{>{\hsize=#1\hsize\centering\arraybackslash}X}%

\begin{document}

\title{An updated dark energy view of inflation}
\date{\today}

\author{Sveva Castello}
\email{sveva.castello@gmail.com}
\affiliation{Department of Physics and Astronomy, Uppsala University, SE-751 20 Uppsala, Sweden}
\author{St\'ephane Ili\'c}
\affiliation{Universit\'e PSL, Observatoire de Paris, Sorbonne Universit\'e, CNRS, LERMA, F-75014, Paris, France}
\affiliation{IRAP, Université de Toulouse, CNRS, CNES, UPS, Toulouse, France}
\author{Martin Kunz}
\affiliation{D\'epartement de Physique Th\'eorique, Universit\'e de Gen\`eve, 1211 Geneva 4, Switzerland}

\begin{abstract}
The present epoch of accelerated cosmic expansion is supposed to be driven by an unknown constituent called dark energy, which in the standard model takes the form of a cosmological constant, characterized by a constant equation of state with $w=-1$. An interesting perspective over the role and nature of dark energy can be achieved by drawing a parallel with a previous epoch of accelerated expansion, inflation, which we assume to be driven by a single scalar field, the inflaton. Since the \emph{Planck} satellite has constrained the value of the scalar spectral index $n_s$ away from 1, the inflaton cannot be identified with a pure cosmological constant, as is also suggested by the fact that inflation ended. Thus, it is interesting to verify whether a hypothetical observer would have been able to measure the deviation of the equation of state parameter of the inflaton from $-1$. To do so, we consider a class of single-field slow-roll inflationary models dubbed HSR$\{i\}$, where the hierarchy of Hubble slow-roll parameters is truncated at the $i$-th order. The models are tested through a Markov Chain Monte Carlo analysis based on combinations of the latest \emph{Planck} and BICEP2/Keck data sets, and the resulting chains are converted into sets of allowed evolution histories of $w$. HSR$\{1\}$ is excluded observationally since it would predict that $n_s=1$, in contrast with the recent \textit{Planck} constraints, while we find that HSR$\{2\}$ would prefer $w > -1$, but is disfavoured by the addition of the BICEP2/Keck data. The overall best description for the data is provided by HSR$\{3\}$, which yields a 68$\%$ upper bound of $1+w < 0.0014$. Therefore, if the current era of accelerated expansion happens to have the same equation of state as inflation during the observable epoch, then current and upcoming cosmological observations will not be able to detect that $w \neq -1$. This provides a cautionary tale for drawing conclusions about the nature of dark energy on the basis of the non-observation of a deviation from $w=-1$.
\end{abstract}

\maketitle

\section{Introduction}

The observational evidence for the accelerated expansion of the Universe  \cite{riess1998observational, perlmutter1999measurements} led to postulating the existence of a cosmic source with a negative pressure, the so-called dark energy. In the framework of the standard cosmological model, dark energy takes the form of a ``cosmological constant", with symbol $\Lambda$, which is interpreted as a vacuum energy with a homogeneous distribution in time and space and is characterized by a constant equation of state $w\equiv \bar{p}/\bar{\rho} =-1$. While its abundance at the present time is well-constrained by observations, yielding approximately 68.9$\%$ of the total cosmic energy density \cite{Planck2018parameters}, its nature and properties are poorly understood, making it one of the key issues in modern theoretical physics.

As already suggested in a previous work \cite{ilic2010dark}, an innovative approach to investigate the role of dark energy consists in drawing a comparison with a postulated earlier epoch of accelerated expansion in the history of the Universe: cosmic inflation \cite{liddle2000cosmological}. Inflation is thought to have led to a rapid increase of the cosmic scale factor some time before Big Bang nucleosynthesis and, in the simplest description, to have been driven by a single scalar field, often dubbed inflaton. While the inflaton can indeed be interpreted as a form of dynamical dark energy, it cannot be identified with a pure cosmological constant with $w=-1$, since it is thought to have rapidly decayed away at the end of inflation. Thus, the question arises whether the equation of state parameter $w$ significantly deviated from -1 during the inflationary era, allowing a hypothetical observer to appreciate the difference with respect to a pure cosmological constant.

In order to constrain the evolution of $w$ during inflation, it is necessary to link it to an observable. The temperature and polarization anisotropies in the cosmic microwave background (CMB) radiation turn out to be ideal candidates, since they are thought to mirror the primordial curvature perturbations generated in the inflationary epoch. In recent years, these anisotropies were characterized with high precision, among others, by the \emph{Planck} mission \citep{2019arXiv190712875P}, yielding strong cosmological constraints from the temperature and polarization maps of the CMB. In this study, we carry out a Monte Carlo Markov Chain analysis with the \emph{Planck} likelihoods from the final 2018 data release \cite{2019arXiv190712875P, 2018arXiv180706210P} and the joint  BICEP2/Keck-WMAP-\emph{Planck} likelihood of \cite{2018PhRvL.121v1301B}, which we expect to provide complementary information. Overall, we aim to update the results obtained in \cite{ilic2010dark} with the most recent  data sets and, additionally, we choose to broaden the previous analysis by distinguishing among three single-field inflationary models with the common assumption of slow roll. The constraints on $w$ and the cosmological implications will be discussed in each case and the best-fitting model will be identified on the basis of Bayesian model selection \cite{trotta2017bayesian}.

\section{Theoretical framework \label{sec:theory}}

\subsection{The equation of state of the inflaton\label{sec:eos}}
Throughout our analysis, we assume that inflation was driven by a single scalar field, the inflaton $\phi$, and that it lasted for longer than approximately 60 e-foldings. This allows us to consider the Universe as spatially flat and to neglect any contribution to the total energy density $\rho$ other than that coming from the inflaton itself. Under these assumptions, we adopted the modelling proposed in \cite{ilic2010dark} in order to relate the equation of state parameter $w$ of the inflaton to standard inflationary quantities. As already pointed out in \cite{ilic2010dark}, this can be achieved by considering the Hubble parameter $H \equiv \dot{a}/a$ as the reference quantity instead of the inflaton potential $V(\phi)$. Since the total energy density is dominated by the inflaton, this allows to directly link the equation of state parameter $w$ to the expansion rate, instead of expressing it in a less convenient form in terms of the pressure and energy density of the inflaton. Through this approach, the equations of motion yield the following expression:
\begin{equation} \label{w_fn_H}
    1+w = - \frac{2}{3} \frac{\dot{H}}{H^2},
\end{equation}
where dots indicate (cosmic) time derivatives. This relation can be further rewritten by introducing a hierarchy of ``Hubble slow-roll" (HSR) parameters $\xi_n$ within the Hamilton-Jacobi formalism \cite{salopek1990nonlinear}, under the standard assumption of slow roll \cite{liddle2000cosmological}. We employ the hierarchy that was originally introduced in \cite{Liddle:1994dx} and choose to adopt the same notation as in \cite{lesgourgues2008wmap}, where the first three parameters are written as
\begin{equation} \label{epsilonH}
    \xi_1 (\phi) \equiv \epsilon_H (\phi) = 2 \mpl \left(\frac{H'(\phi)}{H(\phi)} \right)^2,
\end{equation}
\begin{equation} \label{etaH}
    \xi_2 (\phi) \equiv \eta_H (\phi) = 2 \mpl \, \frac{H''(\phi)}{H(\phi)},
\end{equation}
and
\begin{equation}
    \xi_3 (\phi) \equiv \zeta_H (\phi) = 4 M_\mathrm{Pl}^4  \, \frac{H'''(\phi)}{H(\phi)}  \, \frac{H'(\phi)}{H(\phi)}.
\end{equation}
Here, primes denote derivatives with respect to $\phi$, $M_\mathrm{Pl} \equiv 1/\sqrt{8\pi G}$ is the reduced Planck mass, with $G$ being Newton's constant, and we have set $c = \hbar = 1$ ($G$ is set to one as well further in our analysis). We note that each successive parameter in the hierarchy contains a derivative of $H$ one order above the previous one.

By employing one of the results of the Hamilton-Jacobi formalism,
\begin{equation} \label{Hamilton_Jacobi}
    \frac{\dot{\phi}}{2} = - \mpl \, H' \, ,
\end{equation}
together with Equations~(\ref{w_fn_H}) and~(\ref{epsilonH}) and $H' = \dot{H}/\dot{\phi}$, we obtain
\begin{equation} \label{fundamental_w}
    1+w = \frac{2}{3} \xi_1 \, ,
\end{equation}
which provides the crucial link between $w$ and a standard inflationary quantity. Additionally, it can be shown that the tensor-to-scalar ratio $r$ and the scalar spectral index $n_s$ can also be written in terms of the first two HSR parameters up to the lowest order in slow roll (see e.g. \cite{Liddle:1992wi}):
\begin{equation} \label{r}
    r = 16 \, \xi_1
\end{equation}
and
\begin{equation} \label{n_s}
    n_s - 1 = 2 \xi_2 - 4 \xi_1 \, .
\end{equation}
An interesting conclusion that can be a priori drawn from Equation~(\ref{n_s}) is that, since the \emph{Planck} results have constrained the value of $n_s$ away from 1 at the 8$\sigma$ level \cite{Planck2018parameters}, either $\xi_1$ or $\xi_2$ must be non-zero. Thus, Equation~(\ref{fundamental_w}) and the definition of the HSR parameters imply that we must require that either $w \neq -1$ or $\frac{dw}{dt} \neq 0$ during inflation, in both cases ruling out a pure cosmological constant with $w = \mathrm{const.} = -1$.

\subsection{Inflationary models \label{sec:models}}

The evolution of $w$, given by Equation~(\ref{fundamental_w}), was studied in the context of three single-field inflationary models, with the common underlying assumption of slow roll. Following \cite{lesgourgues2008wmap}, we considered a Taylor expansion of the Hubble parameter $H$ around an arbitrary pivot value of the inflaton $\phi_*$:
\begin{equation} \label{Taylor_H}
    H (\phi- \phi_*) = \sum_{i=0}^{n} \hat{H}_i \, (\phi-\phi_*)^n
\end{equation}
The Taylor coefficient of $n$th order can be expressed in terms of the first $n$ HSR parameters evaluated at $\phi_*$ and an additional parameter, which we denote as $\xi_0^*$:
\begin{equation} \label{xi0}
    \xi_0^* (\phi) \equiv \frac{ H^4 (\phi) }{64 \, H'^2(\phi) \mpl} \biggr\rvert_{\phi_*}.
\end{equation}
We find the following useful relations for the first four Taylor coefficients, where we use the shortened notation $\xi_n^*$ for the HSR parameters evaluated at $\phi_*$:
\begin{equation} \label{Taylor_coeff}
    \begin{split}
        \hat{H}_0 & = \sqrt{\pi \times \xi^*_0 \times \xi^*_1} \\
        \hat{H}_1 & = -\sqrt{4 \pi \times \xi^*_1} \times \hat{H}_0 \\
        \hat{H}_2 & = 4 \pi \times \xi^*_2 \times \hat{H}_0 \\
        \hat{H}_3 & = 4 \pi \times \xi^*_3 \times \frac{\hat{H}_0^2}{\hat{H}_1}.
    \end{split}
\end{equation}
In this study, we chose to truncate the Taylor series for $H(\phi)$ in Equation~(\ref{Taylor_H}) at an order varying between one and three, thus distinguishing between models with two, three and four non-zero HSR parameters respectively \footnote{These truncations are somewhat reminiscent of the flow-equation approach, see e.g.\ \cite{Kinney:2002qn, Hansen:2001eu,Liddle:2003py}.}. The zeroth-order approximation was excluded, since, according to Equation~(\ref{n_s}), a null value of the first two HSR parameters would imply $n_s = 1$ (and additionally $w=-1$ via Equation~(\ref{fundamental_w})), while the latest \emph{Planck} observations \cite{Planck2018parameters} constrain the value of $n_s$ away from 1 by more than 8$\sigma$. In the following, we will employ the labels HSR$\{2\}$, HSR$\{3\}$ and HSR$\{4\}$, such that the HSR$\{n\}$ model corresponds to the case where the $n$th HSR parameter $\xi_n^*$ is the first in the series to be set to zero.

\section{Numerical investigation}
\subsection{Data sets and tools}

In order to constrain the aforementioned models (and their respective parameters), we use the public \emph{Planck} likelihood code and associated data sets\footnote{Available at \url{pla.esac.esa.int}}, namely the latest 2018 release containing the final temperature and polarisation (both E and B modes) measurements from the satellite \cite{2019arXiv190712875P}. We consider both the low- and high-multipole ($\ell$) data as well as the likelihood associated to the lensing convergence map extracted from the same measurements \cite{2018arXiv180706210P}. Additionally, as an alternative to the \emph{Planck} low-multipole B-mode polarisation data, we also consider the joint BICEP2/Keck-WMAP-\emph{Planck} likelihood of \cite{2018PhRvL.121v1301B}, which provides significantly more stringent constraints and should noticeably affect our results. Three data set combinations are considered thereafter:
\begin{itemize}[noitemsep,topsep=1pt]
    \item[(i)] \emph{Planck} low-$\ell$ T/E/B likelihood and high-$\ell$ TT/TE/EE likelihoods (dubbed \emph{P18all} here)
    \item[(ii)] \emph{Planck} low-$\ell$ T/E likelihood, high-$\ell$ TT/TE/EE likelihoods, and low-$\ell$ BICEP2/Keck (\emph{P18$+$BK15})
    \item[(iii)] \emph{Planck} low-$\ell$ T/E likelihood, high-$\ell$ TT/TE/EE likelihoods, low-$\ell$ BICEP2/Keck, and \emph{Planck} lensing likelihood (\emph{P18$+$lens$+$BK15})
\end{itemize}

We investigate the constraints on our models from this choice of data sets using a standard Markov Chain Monte Carlo (MCMC) approach. For this purpose we use ECLAIR, a publicly available\footnote{\url{github.com/s-ilic/ECLAIR}} suite of codes, which interfaces with the popular CLASS Boltzman code \cite{blas2011cosmic} and combines its outputs with likelihoods from state-of-the-art data sets, while using efficient MCMC sampling methods. A detailed description of ECLAIR can be found in the Appendix of \citet{2020arXiv200409572I}, and we only summarize briefly its main features. To sample the parameter space, ECLAIR uses the Goodman-Weare affine-invariant ensemble sampling technique~\citep{GoodmanWeare2010} via the Python implementation \texttt{emcee} \citep{emcee}. The convergence of the MCMC chains is assessed using graphical and numerical tools included in the ECLAIR code package. The resulting chains are then used to determine the marginalized posterior distributions of the parameters using the publicly available Python module \texttt{getdist} \citep{Lewis2019}.

\begin{figure}[h]
    \centering
    \includegraphics[width=\columnwidth]{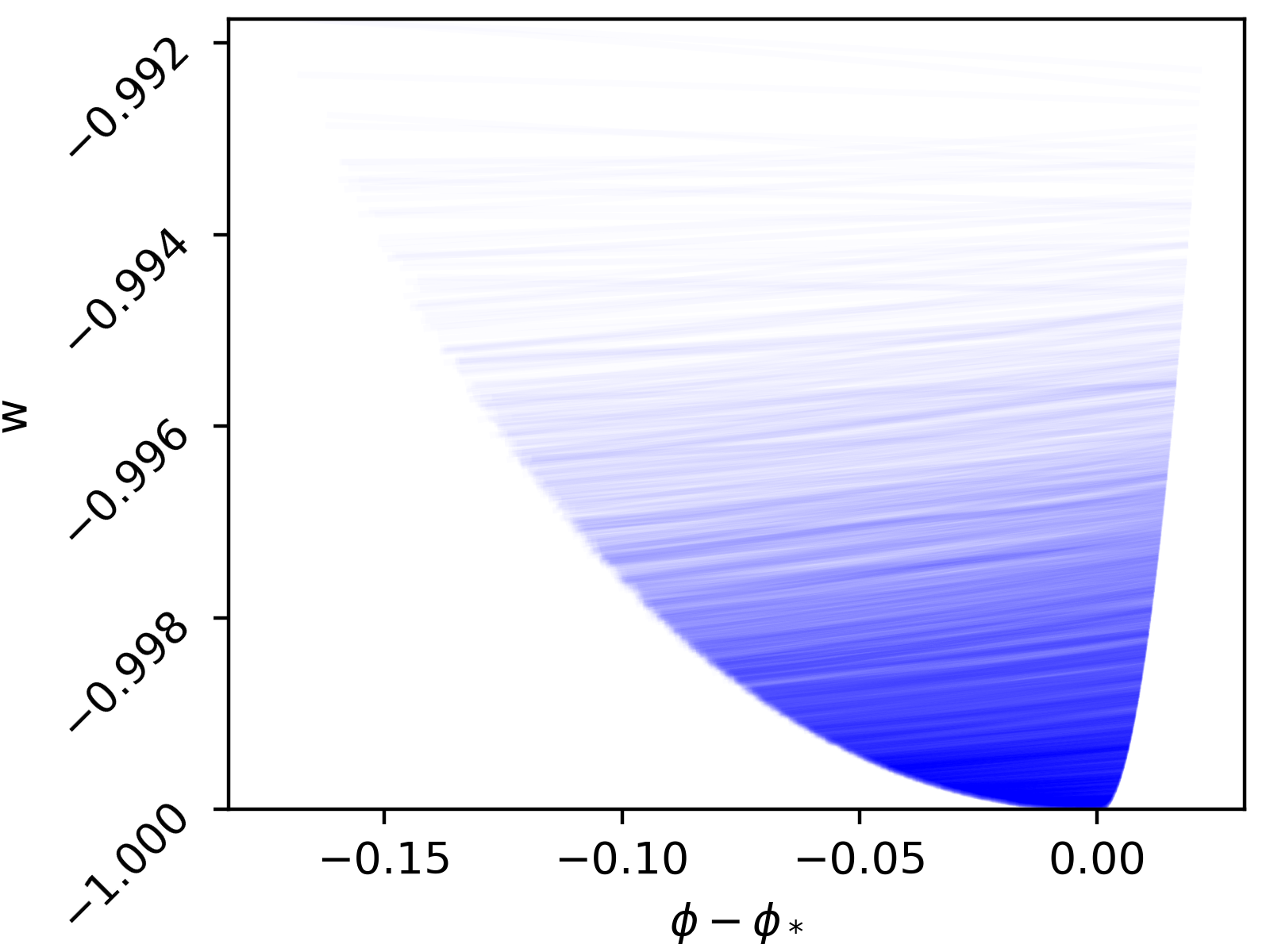}
    \caption{The evolution of the equation of state parameter $w$ as a function of the value of the inflaton field $\phi$ for the HSR$\{3\}$ model and the chains obtained with the \emph{P18all} likelihood. Each line corresponds to the function derived for one sampled point in the parameter space.}
    \label{w_phi_P18_HSR2}
\end{figure}

\begin{figure*}[t]
    \centering
    \includegraphics[width=1.0\linewidth]{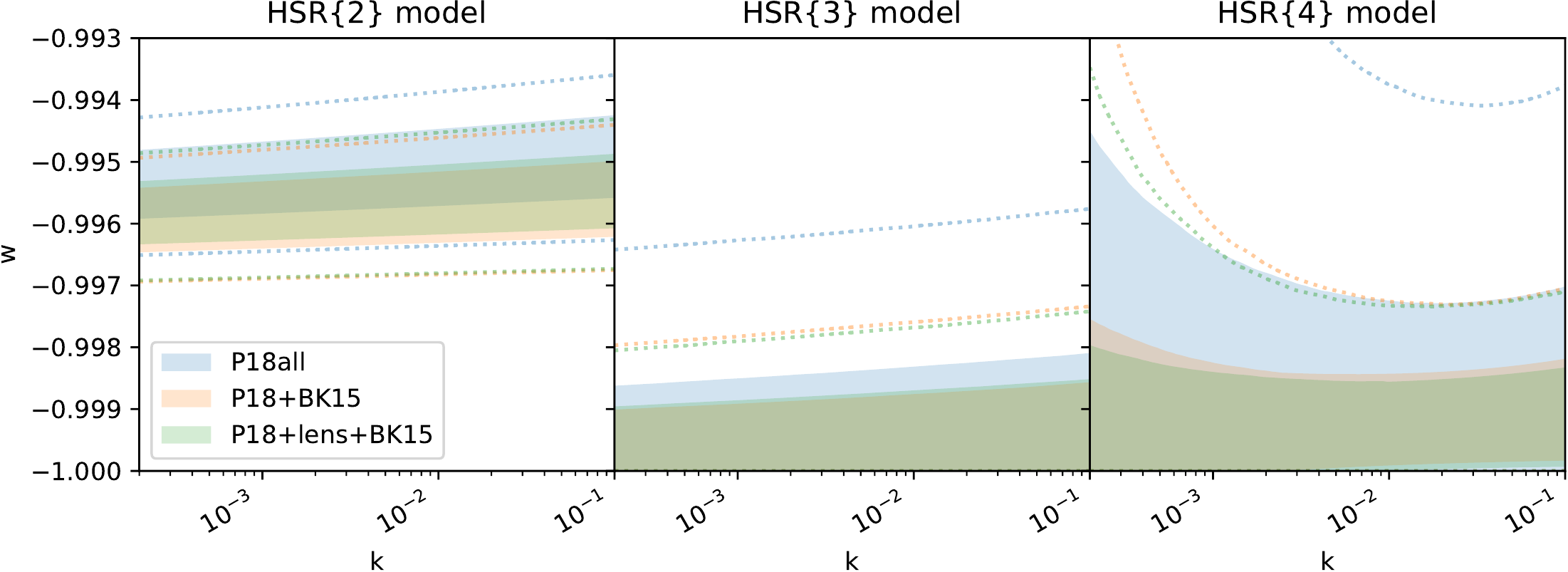}
    \caption{The evolution of $w$ as a function of the comoving wavenumber $k$ at horizon crossing, for each possible pairing of our three tested inflationary models (left, middle and right panels) and three combinations of likelihoods (shown in blue, orange and green). The shaded areas correspond to the 68$\%$ confidence intervals and the dotted lines indicate the 95$\%$ ones.}
    \label{w_k_summary}
\end{figure*}

\subsection{Post-processing} \label{sec:post_processing}

For each tested model, the MCMC chains for the non-zero $\xi^*$ parameters were fed as inputs to an independent Python code computing the equation of state parameter $w$ as a function of the value of the inflaton $\phi$ and the wavenumber $k$ of CMB perturbations. In order to achieve this, the expression for $H(\phi)$ was reconstructed according to Equations~(\ref{Taylor_H}) and (\ref{Taylor_coeff}) and was then used to derive $w(\phi)$ from Equations~(\ref{epsilonH}) and (\ref{fundamental_w}). As an example, Figure~\ref{w_phi_P18_HSR2} shows the set of extrapolated $w(\phi)$ functions obtained for the HSR$\{3\}$ model with the \emph{P18all} likelihood. The range of $\phi$ values varies for each choice of parameter values and decreases as $w$ becomes more negative. The physical interpretation of this behaviour is that the inflaton rolls more and more slowly along the shape of the potential, until it ``freezes" for $w = -1$. This is due to the fact that, according to Equation~(\ref{fundamental_w}), the more $w$ approaches $-1$, the closer $\xi_1$ will be to zero, yielding a more and more extreme slow-roll regime.

This feature makes the constraints on $w$ rather difficult to interpret from the $w(\phi)$ plots. A clearer representation can instead be obtained in $k$ space, as shown in Figure~\ref{w_k_summary} for all combinations of tested models and likelihoods. The $w(k)$ functions can be obtained by associating a $k$ value to a given $\phi$ through the ``horizon-crossing condition" \cite{baumann2009tasi}, yielding the (comoving) scale at which the CMB perturbations cross the Hubble radius and thus become cosmologically relevant:
\begin{equation} \label{k(a)}
    k = a H(\phi(a)),
\end{equation}
where $a$ is the scale factor.

These computations require to select an interval of scales of interest. Figure~17 from \cite{Planck2018inflation} suggests that the range that is accessible to observations by the \emph{Planck} satellite approximately lies between $10^{-3}$ and $10^{-1}$ Mpc$^{-1}$. Thus, we adopted the interval proposed by \cite{lesgourgues2008wmap}, $[k_{\rm min}, k_{\rm max}] = [2 \times 10^{-4}, 0.1]$ Mpc$^{-1}$, which encloses the observable range and also allows us to test the overall behaviour right outside of it. The pivot scale for the primordial power spectrum was set to the standard value in CLASS, $k_* = 0.05$ Mpc$^{-1}$. On the other hand, the range of tested $\phi$ values can be derived by obtaining an expression for $\frac{da}{dk}$ from Equation~(\ref{k(a)}) and integrating $\frac{d\phi}{dk} = \frac{d \phi}{da} \frac{da}{dk}$ over the chosen $k$ range. Thus, while the interval in $k$ is fixed, the one in $\phi$ depends on the reconstructed functional form of $H$ and, consequently, on the sampled point in the parameter space.

From the values of the $\mathrm{HSR}$ parameters for each entry in the MCMC chains, we also obtain the primordial scalar spectral amplitude $A_s$, the scalar spectral index $n_s$ and the scalar-to-tensor ratio $r$ as outputs of the CLASS Boltzmann code. For our models, we find that these outputs agree with the expression for $r$ in Equation~(\ref{r}), while the one for $n_s-1$ in Equation~(\ref{n_s}) has small corrections for HSR$\{4\}$ due to higher-order contributions.

\subsection{Constraints from the MCMC analysis}

The reconstructed constraints on $w(k)$ obtained for the three tested models are shown in Figure~\ref{w_k_summary} and listed in Table~\ref{tab:wbounds}. The full one- and two-dimensional posterior distributions on model (and derived) parameters are given in Figure~\ref{triangle_plot} in the Appendix.

We immediately notice a strong dependence of the results on the model choice: for HSR$\{2\}$, we detect a clear deviation from $w=-1$, while the $w$-posterior for HSR$\{3\}$ and HSR$\{4\}$ are compatible with $w=-1$. Moreover, the results for HSR$\{4\}$ show a flaring-up of the errors towards the limits of the chosen $k$-range. In the following, we will discuss these features in turn.

\subsubsection{HSR$\{2\}$ model}
The detection of $w \neq -1$ in the HSR$\{2\}$ model can be interpreted as being mainly related to the equilibrium between the preferred values of the scalar spectral index $n_s$ and the tensor-to-scalar ratio $r$. In fact, since $\xi_2$ is zero in this model, a non-trivial value of $\xi_1$ is solely responsible for pushing $n_s$ away from 1 in Equation~(\ref{n_s}), as required by observations. This also yields $r \neq 0$ according to Equation~(\ref{r}), which would imply non-zero tensor perturbations. While this is not a priori impossible, no unambiguous evidence for gravitational waves has ever been found in previous studies, suggesting that the data would instead prefer smaller values of $r$. This results in opposite effects pushing $r$ both towards and away from zero at the same time and an equilibrium is reached for a value of $n_s$ slightly closer to 1 than in the other models and a non-zero $\xi^*_1$.

\subsubsection{HSR$\{3\}$ model}

When allowing $\xi^*_2$ to vary in the HSR$\{3\}$ model, $n_s$ and $r$ (equivalently $w$) are no longer strongly correlated. In this case $\xi^*_1$, which governs the $r$ and $w$ constraints, is able to reach zero, which is the value preferred by the data according to the marginalized constraints. On the other hand, $\xi^*_2$ is correlated with $\xi^*_1$, so that in particular the linear combination $\xi^*_2 - 2 \xi^*_1$, which is proportional to $n_s-1$ according to Equation~(\ref{n_s}), is always non-zero.

The fact that the posterior shifts strongly when a new parameter is added is an indication that the more complex model HSR$\{3\}$ is preferred. As we will discuss below, this is however not a sufficient condition when using Bayesian model comparison due to the ``Occam's razor'' factor inherent in this approach.

Nonetheless, based on the model comparison performed in Section~\ref{sec:model_comparison}, we consider the HSR$\{3\}$ model the most appropriate choice within our model classes. We note that the resulting 68$\%$ upper bound of $1+w < 0.0014$ significantly improves the result obtained in the previous study from \cite{ilic2010dark}.

\subsubsection{HSR$\{4\}$ model}

The resulting bound on $w$ is similar to the HSR$\{3\}$ model but weaker, reflecting the addition of a free parameter. In this case, however, the value of $w$ blows up at the boundaries of the chosen $k$ interval, as can be noticed from Figure~\ref{w_k_summary}. This result indicates that only a subset of the interval is constrained by the data, an interpretation that is consistent with the constraints on the observable \emph{Planck} scales from \cite{Planck2018inflation}. Moreover, the posterior distributions in Figure~\ref{triangle_plot} show that setting $\xi^*_3$ to zero is compatible with the constraints, suggesting that this model is disfavoured. Nevertheless, it is again necessary to turn to model comparison for a more quantitative check.

\begin{table}
\caption{95\% confidence limits on $1+w$ for all combinations of the different models and likelihood sets considered.}
\label{tab:wbounds}

\begin{tabularx}{\columnwidth}{L{1.3} C{0.9} C{0.9} C{0.9}}
 & HSR$\{2\}$ & HSR$\{3\}$ & HSR$\{4\}$ \\
\hline
\emph{P18all}  & 0.0050 $\pm$0.0013   & $<$0.0042 & $<$0.0059 \\
\emph{P18+BK15}  &  0.0044 $\pm$0.0012 & $<$0.0026 & $<$0.0028 \\
\emph{P18+lens+BK15} & 0.0045 $\pm$0.0012 &  $<$0.0025 & $<$0.0029 \\
\hline
\end{tabularx}
\end{table}

\subsection{Bayesian model comparison} \label{sec:model_comparison}
In order to quantitatively identify the model providing the `best' description, we employ the tools of Bayesian model probability. In particular, we compute the Bayes factor, which, given two models $M_1$ and $M_2$, is defined as the ratio of marginal likelihoods:
\be
B_{12} = \frac{P(D|M_1)}{P(D|M_2)} \, .
\ee
Here, $P(D|M_i)$ is the probability of observing data $D$ in the model $M_i$. If all models are equally likely \emph{a priori}, this ratio also corresponds to the relative model probability, such that model $M_1$ is favoured for $B_{12}>1$, while $M_2$ is preferred in the opposite case. We underline that the Bayes factor should be seen as `betting odds' and can be interpreted according to Jeffreys' scale \cite{Jeffreys:1939xee}, which roughly states that $|\log_{10} B_{12}|>1$ can be considered as strong evidence and $|\log_{10} B_{12}|>2$ as decisive.

In general, the quantity $B_{12}$ is rather difficult to obtain. However, the computations are simplified in the special case of nested models, i.e. when a more complex model $M_1$ with (for simplicity) one extra parameter $p$ becomes equivalent to a simpler model $M_2$ when $p$ is set to a specific value $p_*$. In this case, the Bayes factor can be obtained through a procedure often dubbed Savage-Dickey Density Ratio (SDDR, see e.g.\ \cite{Trotta:2005ar}):
for a common parameter vector $\bq$, the Bayes factor between the two models is simply the ratio of the prior to the posterior for the parameter $p$ at the nested point, marginalized over the common parameters $\bq$,
\be
B_{12} = \frac{P(p=p_*|M_1)}{P(p=p_*|M_1,D)} \, . \label{eq:sddr}
\ee
The SDDR makes a key point of Bayesian model comparison quite explicit: if the probability of $p=p_*$ in the posterior is very small, then the more complicated model will be favoured, as in the usual relative goodness of fit comparison between the two models. However, even if the simpler model provides a slightly worse fit, it can receive a significant boost from the numerator of Equation~(\ref{eq:sddr}) if the prior is much wider than the posterior. This is often called the ``Occam's razor'' factor.

In our analysis, the model HSR$\{$n$\}$ is always nested in the model HSR$\{$n+1$\}$ and corresponds to the case where the parameter $\xi^*_n$ is set to zero. Therefore, we can always employ the SDDR in order to compute the Bayes factor, which requires to select a prior on the nesting parameter $\xi^*_n$. Since we would naturally expect the slow-roll parameters to be smaller than, but generally of order 1, a possible choice is to use a flat prior on the range $[-1,1]$. However, since we also expect these parameters to be `small' during slow roll, it makes sense to also consider narrower priors in the form $[-A,A]$, where $A\ll 1$. In this study, we choose to adopt $A\approx |n_s-1|\approx 0.04$, since the deviation from a scale-invariant spectrum provides an observational characterization of the order of the slow-roll parameters during the relevant period for our analysis. According to the SDDR in Equation~(\ref{eq:sddr}), it is obvious that the Bayes factor in favour of the simpler model for a prior with width $2 A$ is equal to $1/A$ times the Bayes factor of width $2$, such that the narrower prior given above makes the simpler model $25$ times less favoured relative to the wide prior.

Table~\ref{tab:models} contains the Bayes factors for all HSR models for the selected flat priors widths $[-1,1]$ (the `wide prior' case) and $[-0.04,0.04]$  (the `SR prior case'), always relative to the simplest model, HSR$\{2\}$. We also give the relative best-fit $\chi^2$ value for the models, even though we underline that the $\chi^2$ difference is in general not a good model comparison quantity, since a more complicated nested model necessarily always has a lower minimal $\chi^2$ (i.e.\ it corresponds to the denominator part of Equation~(\ref{eq:sddr}) only).

\begin{table}
\caption{Bayes factors $B_{\{i\}\{2\}}$ and $\Delta\chi^2$ values between the different HSR$\{i\}$ models considered here, always relative to the HSR$\{2\}$ model. The `wide prior' case corresponds to a prior of $[-1,1]$ on all slow-roll parameters, and the `SR prior' case to a prior of $[-0.04,0.04]$.}\label{tab:models} 

\begin{tabularx}{\columnwidth}{L{1.3} C{1} C{1} C{0.7}}
Model / \emph{data}  & $B_{\{i\}\{2\}}$ [wide]  & $B_{\{i\}\{2\}}$ [SR] & $\Delta\chi^2$ \\
\hline
\emph{P18all} &&& \\
HSR$\{2\}$   & $1$   &  $1$ &   $0$  \\
HSR$\{3\}$   & $0.090$  &  $2.3$  & $-5.9$ \\
HSR$\{4\}$  & $0.011$  & $2.3$   & $-6.5$ \\
\hline
\emph{P18+BK15} &&& \\
HSR$\{2\}$   & $1$   &  $1$ &   $0$  \\
HSR$\{3\}$  & $1.3$  &  $31$  & $-13.4$ \\
HSR$\{4\}$   & $0.14$  & $33$   & $-14.0$ \\
\hline
\emph{P18+lens+BK15} &&& \\
HSR$\{2\}$  & $1$ &   $1$  & $0$  \\
HSR$\{3\}$  & $2.4$  &  $61$  & $-13.8$ \\
HSR$\{4\}$   & $0.22$  & $62$  & $-14.0$ \\
\hline
\end{tabularx}
\end{table}

\subsubsection{HSR$\{2\}$ vs HSR$\{3\}$}

When considering the \emph{P18all} likelihood, the normalised posterior for HSR$\{3\}$ at $\xi_2^*=0$ (i.e.\ the point where HSR$\{3\}$ is nested in HSR$\{2\}$) marginalized over all other parameters is
\be
P(\xi_2^* = 0) = 5.55  \qquad \text{for \emph{P18all}} \, .
\ee
The normalized prior on $\xi_2^*$ is equal to $0.5$ (wide prior) or $12.5$ (SR prior). Therefore, this results in a Bayes factor of $11.1$ in favour of HSR$\{2\}$ (the simpler model) for the wide prior, and of $2.3$ in favour of HSR$\{3\}$ for the SR prior, indicating either a strong preference for the simpler model or an effectively undecided outcome, depending on the prior width.

The fact that the HSR$\{2\}$ model is preferred over the HSR$\{3\}$ model for the wide prior may appear surprising, given that the HSR$\{3\}$ fits the data better, with a $\Delta\chi^2=-5.9$ for a single extra parameter. As mentioned above, this is due to the ``Occam's razor'' factor: the 95\% constraint on $\xi_2^*$ in HSR$\{3\}$ is
\be
\xi_2^* = -0.013^{+0.010}_{-0.008} \qquad \text{for \emph{P18all}},
\ee
which is much narrower than the priors, in particular than the wide prior. This significant shrinking of the parameter space into a region `close' to the simpler model prediction boosts the relative probability of HSR$\{2\}$ and makes it competitive with the HSR$\{3\}$ model. This is particularly interesting because HSR$\{2\}$ gives radically different results from the other models, yielding $w \neq -1$ and consequently leading to a `detection' of primordial gravitational waves.

A solution to this \emph{impasse} consists in including additional trustworthy data sets that are compatible with the already used ones. In the \emph{P18+BK15} data set, we chose to add the low-$\ell$ BICEP2/Keck data that constrains the $B$-modes of the CMB polarisation much better. We additionally considered the \emph{P18+lens+BK15} data set, with the inclusion of the \emph{Planck} lensing likelihood, but we find that the results obtained in this case are qualitatively similar to \emph{P18+BK15}.

Table~\ref{tab:models} clearly shows that the $\chi^2$ difference between HSR$\{2\}$ and HSR$\{3\}$ is much larger for \emph{P18+BK15}, suggesting that the goodness of fit component of the SDDR will now indeed favour HSR$\{3\}$ more. Indeed, we find that
\be
P(\xi_2^* = 0) = 0.40  \qquad \text{for \emph{P18+BK15}} \, .
\ee
The resulting Bayes factors are then $1.3$ (wide prior) and $31$ (SR prior), both in favour of HSR$\{3\}$. While the choice of wide prior still does not lead to a significant preference for HSR$\{3\}$, the SR prior choice now strongly favours HSR$\{3\}$ over HSR$\{2\}$ according to Jeffreys' scale. The addition of the \emph{Planck} lensing data strengthens this preference by about a factor of two and, while this does not qualitatively change the outcome, it reinforces our view that HSR$\{3\}$ should be preferred over HSR$\{2\}$. Additionally, HSR$\{3\}$ could be considered more natural than HSR$\{2\}$, as both $\xi_1$ and $\xi_2$ contribute to $n_s$ at leading order in slow-roll according to Eq.\ (\ref{n_s}).

\subsubsection{HSR$\{4\}$ and models with more parameters}

We again begin by considering the \emph{P18all} data set, yielding
\be
P(\xi_3^* = 0) = 4.0  \qquad \text{for \emph{P18all}} \,
\ee
for the normalised posterior at the point where HSR$\{3\}$ is nested in HSR$\{4\}$.
This looks comparable to the result in the previous subsection, but the situation is actually somewhat different. In fact, the 95\% confidence bounds on the extra parameter are now
\be
\xi_3^* = 0.06\pm 0.09 \qquad \text{for \emph{P18all}} \, ,
\ee
i.e. the value of $\xi_3^*$ is well compatible with zero, but the error bars are wider, such that the value of the normalised posterior at the peak is lowered with respect to above. Including further data sets does not significantly change the situation. This reflects another property of the Bayes factor that can be well understood from the SDDR: if we add a completely unconstrained parameter (for example one which the problem at hand simply does not depend on), the posterior will be equal to the prior. In that case, Equation~(\ref{eq:sddr}) implies that $B_{12}=1$, i.e.\ the Bayes factor does not distinguish between the two models and, in general, the simpler model should be taken as the preferred one. Therefore, even though the HSR$\{3\}$ model is only favoured by a factor of $8$ with the wide prior and yields $B_{\{4\}\{3\}}=1.0$ with the SR prior for the reason explained above, we consider HSR$\{3\}$ as the model providing the `best' description to the data.

We can also compute the Bayes factor between HSR$\{4\}$ and HSR$\{2\}$ through the relative model probabilities
\be
\frac{P(D|\mathrm{HSR}\{4\})}{P(D|\mathrm{HSR}\{2\})} = \frac{P(D|\mathrm{HSR}\{4\})}{P(D|\mathrm{HSR}\{3\})}  \frac{P(D|\mathrm{HSR}\{3\})}{P(D|\mathrm{HSR}\{2\})} \, .
\ee
As HSR$\{4\}$ is never significantly preferred over HSR$\{3\}$, this leads again to a strong (nearly decisive) preference for HSR$\{2\}$ over HSR$\{4\}$ for the wide prior choice, and an undecided outcome for the SR prior when only considering the \emph{P18all} data. For \emph{P18+BK15}, we obtain effectively the same outcome as for HSR$\{3\}$, i.e.\ no strong indication is found with the wide prior, while HSR$\{2\}$ is disfavoured with the SR prior.

When truncating the HSR hierarchy at higher order in models HSR$\{5\}$, HSR$\{6\}$, and so on, the additional parameters will in general be even more weakly constrained than $\xi_3^*$. Since already HSR$\{4\}$ is not preferred over HSR$\{3\}$ precisely because of the weak constraint on the extra parameter, it is quite unlikely that including further parameters will provide a better description of the data. For this reason, we do not investigate models that involve a higher-order expansion than HSR$\{4\}$.

\subsection{A comment on the standard cosmological model}

When studying the cosmological standard model at late times,  the only inflation-related parameters that are usually varied are $A_s$ and $n_s$, while $r$ is generally assumed to be zero. In the context of single-field slow-roll inflation models, this is not exactly possible, as all light degrees of freedom, including gravitons, are excited during the period of accelerated expansion. For this reason, we are also not able to set $r$ exactly to zero, as this would require setting $\xi_1^*=0$, which would result in all Taylor coefficients vanishing (for finite $\xi_0^*$), cf.~Equation~(\ref{Taylor_coeff}).

We can however simulate this situation by choosing $\xi_1^*$ to be very small \emph{a priori}. As $d \ln(\xi_1)/d\ln a = 2 (\xi_2-\xi_1)\approx n_s-1$ when $\xi_1\ll \xi_2$ (see e.g.\ \cite{ilic2010dark}), we see that $\xi_1$ will remain small for many e-foldings if it is set to be small enough initially. This is therefore not an impossible model, but it appears rather unnatural to have $\hat{H}_1 \ll \hat{H}_2$ in the Taylor expansion about the arbitrary pivot value $\phi_*$.  Nonetheless, it is possible to make $1+w$ arbitrarily small in this way, yielding an agreement with observations as good as in the HSR$\{3\}$ model, except that $\xi_1^*$ is limited to tiny values through its prior.

\section{Conclusions}

In this paper, we revisit the results of \cite{ilic2010dark} concerning the bounds on the equation of state parameter of the inflaton. The original constraints were obtained using the CMB measurements of the WMAP satellite, and we find that the \emph{Planck} satellite data, especially when combined with the BICEP2/Keck data, reduces the uncertainty on the equation of state parameter $w$ by about one order of magnitude.

We choose to describe (single-field) inflation by introducing a hierarchy of Hubble slow-roll parameters. Within this formalism, we consider three different models corresponding to truncating the hierarchy, and therefore a Taylor expansion of the Hubble parameter $H$ during inflation, at order 2, 3 or 4. From a pure goodness-of-fit perspective, the models with 3 or 4 parameters provide a better fit to the data than the simplest model with only two free parameters. However, Bayesian model comparison, which includes an ``Occam's razor'' factor based on the shrinking of the parameter space between prior and posterior, indicates that the 2-parameter model is not significantly disfavoured when employing the \emph{Planck} CMB data only. Adding the BICEP2/Keck data sets instead leads to a weak to strong preference for more than two parameters, depending on the prior.

The choice of model has important physical implications in the context of our analysis. In the case of the simplest model, in fact, we find that the value of the equation of state parameter $w$ is directly linked to the deviation $n_s-1$ of the scalar perturbations from a scale-invariant spectrum. Thus, since \emph{Planck} detects $n_s < 1$ with a significance over 8$\sigma$, we obtain a strong constraint of $w>-1$, which also implies a non-zero value of the tensor-to-scalar ratio $r$ and therefore the presence of primordial gravitational waves. On the other hand, this behaviour is not observed in the case of more complex models, where $w=-1$ and $r=0$ are included in the posterior. Thus, the fact that the 2-parameter model is disfavoured when considering additional data sets is a non-trivial result. 

Based on the comparison performed with the combined \emph{Planck} and BICEP2/Keck data sets, we conclude that the best description is provided by the three-parameter model, HSR$\{3\}$, for which we obtain a 68\% upper limit of $1+w < 0.0014$. It is interesting to note that, through Equations~(\ref{Taylor_H}) and (\ref{Taylor_coeff}), the preference for this model indicates that both $H'(\phi_*)$ and $H''(\phi_*)$ are non-zero. This suggests a fairly complex time-evolution of the `primordial dark energy', requiring a description with at least two parameters.

This result provides useful insights when put into relation with the present cosmic epoch. Indeed, while in general there is no direct link between the inflaton dynamics in the early universe and the late-time dark energy,\footnote{It is possible to construct models where the early and late `dark energy' are connected. Since this is somewhat outside the scope of this article, we only mention here the Higgs-Dilaton model \cite{GarciaBellido:2011de,Trashorras:2016azl} where indeed a late-time dark energy equation of state very close to $w=-1$ is predicted.} it is nonetheless interesting to compare the two, as both phenomena lead to a period of accelerated expansion. We underline that this type of comparison is generally complicated, and the translation of our inflation results to today's dark energy requires the hypothesis that the latter happens to be in a regime similar to the inflaton during the period when the observable scales left the horizon, i.e. a slow-roll regime. Provided that this is the case, the resulting deviation of the equation of state from $w=-1$ would be around one order of magnitude smaller than the expected precision of the next generation of cosmological surveys even under optimistic assumptions (see e.g. \cite{2020A&A...642A.191E}), and it would thus be difficult to detect it.

Therefore, the results of this study can be interpreted as a cautionary tale for the ongoing quest for the nature of dark energy. The lack of an observational detection of $w \neq -1$ in the next decade might reinforce the conclusion that the current accelerated expansion of the Universe is indeed driven by a cosmological constant. However, provided that the physical phenomena underlying inflation and the present epoch can be compared, our analysis implies that, if $w$ remains compatible with $-1$, strong conclusions concerning the nature of dark energy will still be premature.

\begin{acknowledgments}
It is a pleasure to thank Andrew Liddle for useful comments on the draft. MK acknowledges funding from the Swiss National Science Foundation.
\end{acknowledgments}

\appendix
\section{Triangle plots from the MCMC computations}
\begin{figure*}[h]
    \includegraphics[width=1.0 \linewidth]{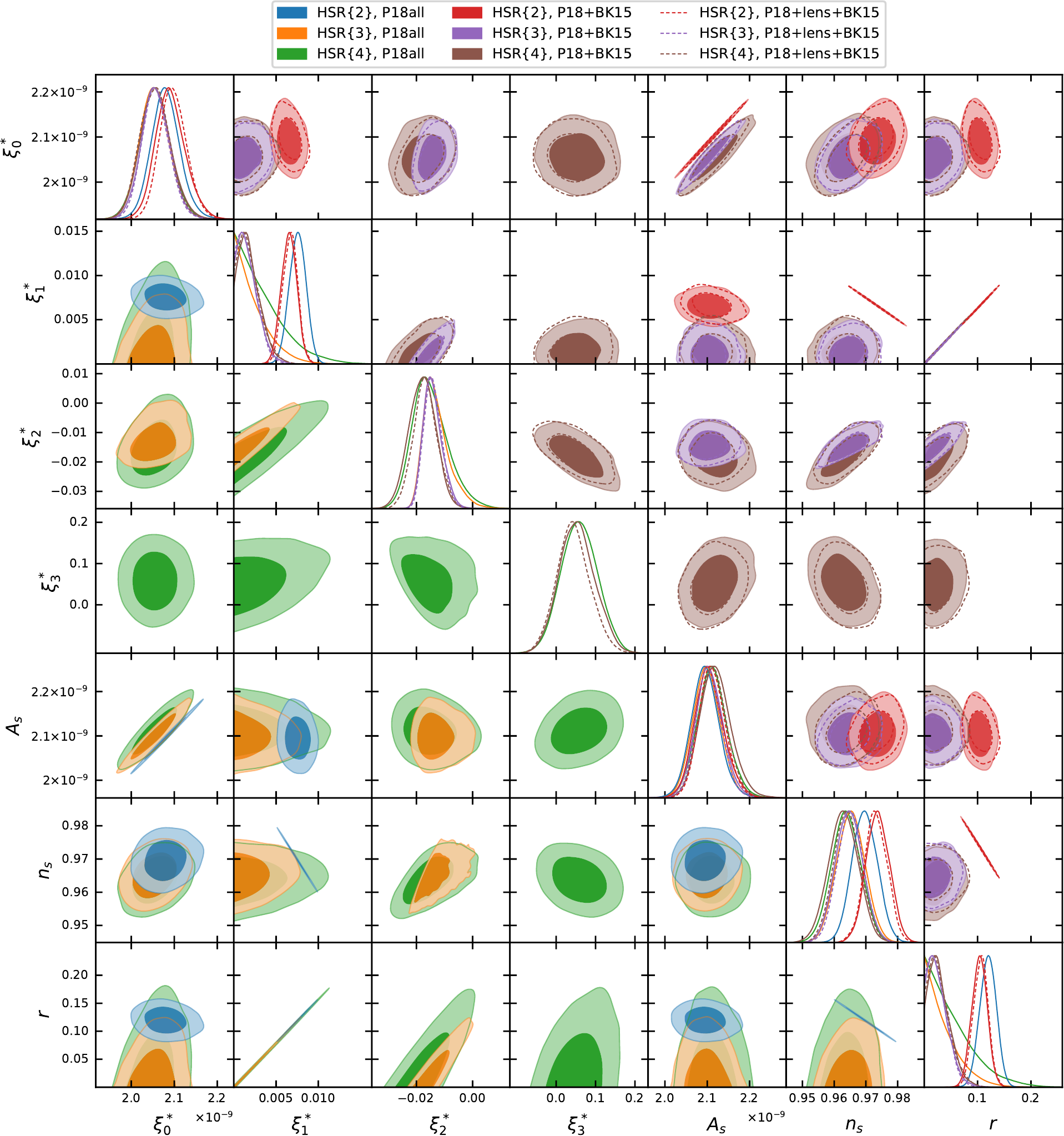}
    \caption{2D confidence contours (68 and 95\%) and 1D marginalised posterior
    distributions for the HSR parameters, the primordial power spectrum amplitude $A_s$, the scalar spectral index $n_s$, and the scalar-to-tensor ratio $r$ obtained with all our combinations of models and data sets. The plots for the HSR$\{2\}$, HSR$\{3\}$ and HSR$\{4\}$ models are shown respectively in blue, orange and green when referring to the results obtained with the \emph{P18all} data set in the lower left triangle of the plot, and in red, violet and brown when including the BICEP2/Keck (solid) and \emph{Planck} lensing data too (dashed) in the upper right part.}
    \label{triangle_plot}
\end{figure*}

In Figure~\ref{triangle_plot}, we show the detailed 1D and 2D marginalised posterior distributions for our model parameters $\xi^*_i$ (with $i=0,1,2,3$) and the derived parameters $A_s$, $n_s$ and $r$ for all combinations of data sets used in the present work.

\bibliography{references.bib}

\end{document}